\documentclass[twocolumn,floatfix,showpacs,superscriptaddress]{revtex4-1}
\usepackage{inputenc}
\usepackage{graphicx,epsfig,amssymb,multirow}

\begin{document}

\title{Anatase TiO$_2$ Nanowires Functionalized by Organic Sensitizers for Solar Cells : \\
A Screened Coulomb Hybrid Density Functional Study}

\author{Hatice \"{U}nal}
\affiliation{Deparment of Physics, Bal{\i}kesir University, Bal{\i}kesir 10145, Turkey}

\author{Deniz Gunceler}
\affiliation{Deparment of Physics, Cornell University, Ithaca, NY 14853, USA}

\author{O\u{g}uz G\"{u}lseren}
\affiliation{Department of Physics, Bilkent University, Ankara 06800, Turkey}

\author{\c{S}inasi Ellialt{\i}o\u{g}lu}
\affiliation{Basic Sciences, TED University, Ankara 06420, Turkey}

\author{Ersen Mete}
\email{emete@balikesir.edu.tr}
\thanks{Corresponding author}
\affiliation{Deparment of Physics, Bal{\i}kesir University, Bal{\i}kesir 10145, Turkey}

\date{\today}

\begin{abstract}
The adsorption of two different organic molecules cyanidin glucoside
(C$_{21}$O$_{11}$H$_{20}$) and TA-St-CA on anatase (101) and (001) nanowires
have been investigated using the standard and the range separated hybrid
density functional theory calculations. The electronic structures and optical
spectra of resulting dye--nanowire combined systems show distinct features
for these types of photochromophores. The lowest unoccupied molecular
orbital of the natural dye cyanidin glucoside is located below the conduction
band of the semiconductor while, in the case of TA-St-CA, it resonates with
the states inside the conduction band. The wide-bandgap anatase nanowires can be
functionalized for solar cells through electron-hole generation and subsequent
charge injection by these dye sensitizers. The intermolecular charge transfer
character of Donor-$\pi$-Acceptor type dye TA-St-CA is substantially modified
by its adsorption on TiO$_2$ surfaces. Cyanidin glucoside exhibits relatively
stronger anchoring on the nanowires through its hydroxyl groups. The atomic
structures of dye--nanowire systems re-optimized with the inclusion of nonlinear
solvation effects showed that the binding strengths of both dyes remain
moderate even in ionic solutions.
\end{abstract}


\maketitle

\section{Introduction}

The wide bandgap metal oxide, TiO$_2$ has gained an increased attention since 
the discovery of its ability to carry out hydrolysis under UV irradiation by 
Fujishima and Honda.\cite{Fujishima1,Diebold,Khan,MChen,WGZhu,Yin,Celik1} In 
addition to showing such an excellent photocatalytic performance, TiO$_2$ has 
also become the material of choice as the anode in dye sensitized solar cell 
(DSSC) applications due to its favorable electrochemical and charge carrier 
conduction properties.\cite{ORegan,Hangfeldt,Gratzel}

In terms of more efficient utilization, titania possesses particularly 
importance with its nanocrystalline 
forms.\cite{Naicker,Boercker,Iacomino,Fuertes} Quasi-one-dimensional TiO$_2$ 
nanostructures offer high surface-to-volume ratios that is desirable to improve 
efficiencies of photovoltaic and photocatalytic 
processes.\cite{XChen,Cakir1,Cakir2}

The (001) and (101) terminations of the anatase polymorph are known to exhibit 
remarkably higher photocatalytic activity relative to the surfaces of the rutile 
phase.\cite{Hengerer,Lazzeri,Thomas,Selloni} Moreover, the naturally 
occurring anatase form has been reported to be the most stable phase of TiO$_2$ 
in nanodimensions.\cite{Naicker,Boercker,Iacomino,Fuertes}

\begin{figure*}[t!]
\epsfig{file=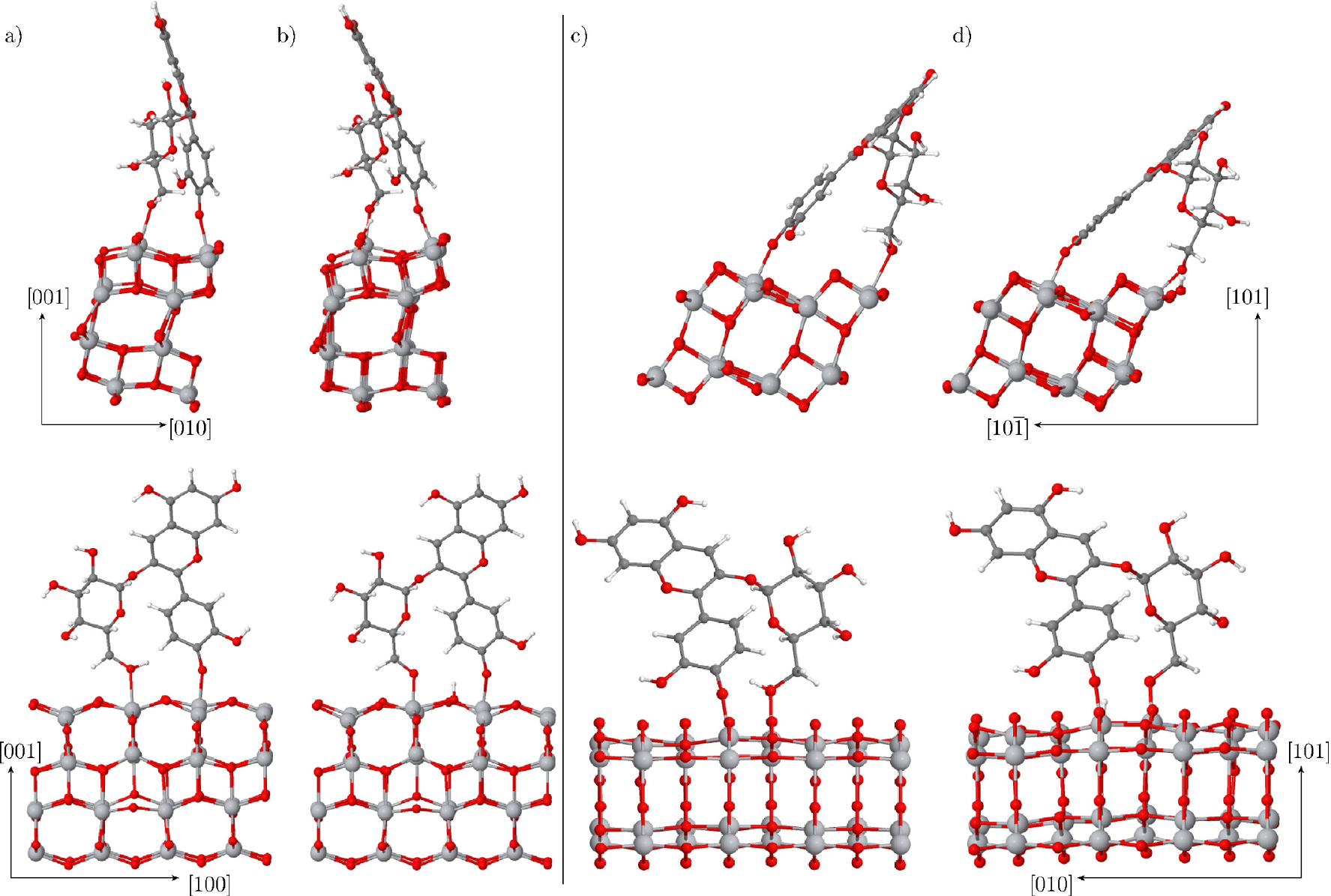,width=17cm}
\caption{Optimized adsorption geometries of cyanidin glucoside on anatase 
(001)-nanowire and (101)-nanowire\label{fig1}}
\end{figure*}

\begin{figure*}[t!]
\epsfig{file=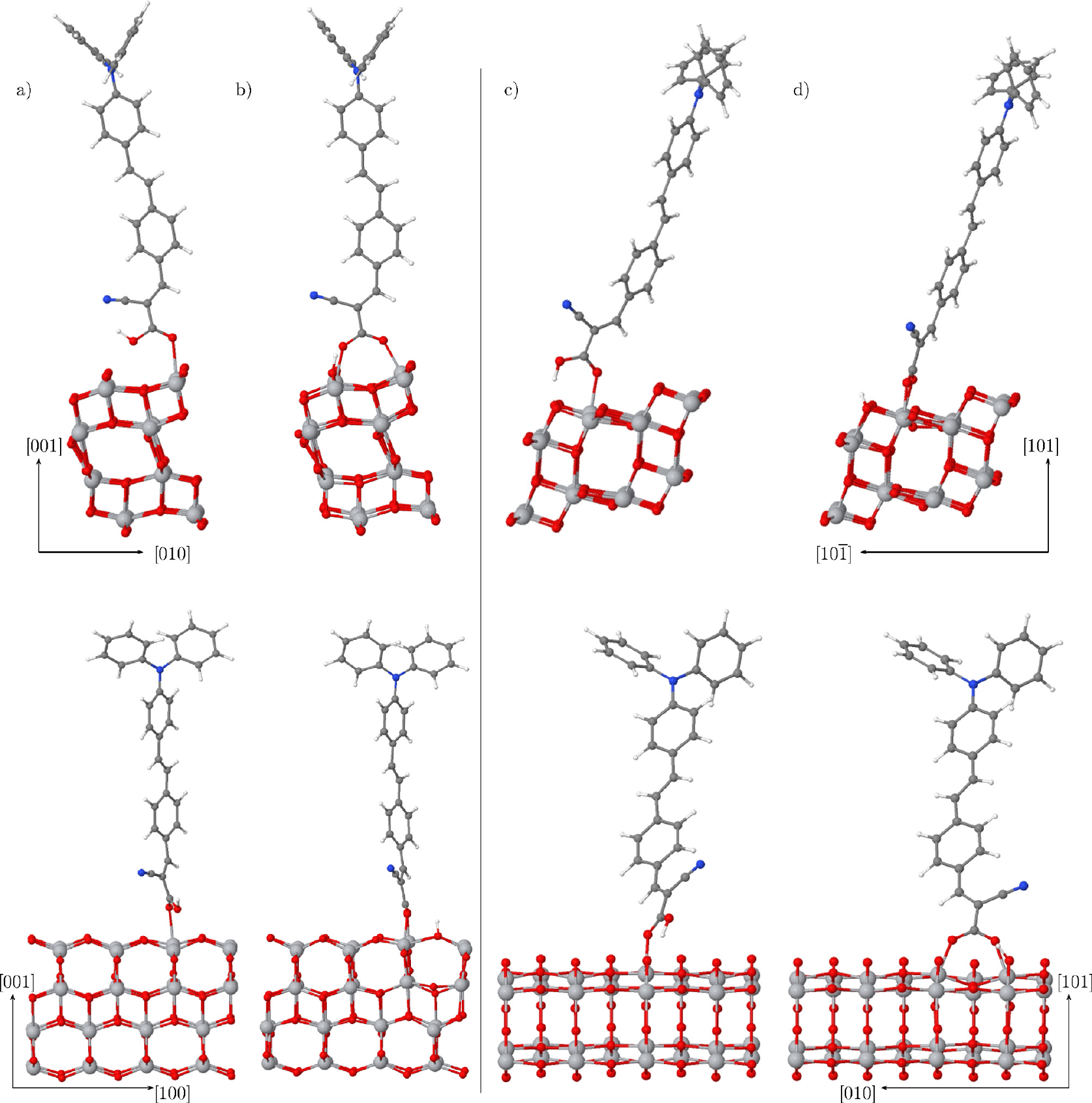,width=17cm}
\caption{Optimized adsorption geometries of TA-St-CA on anatase (001)-nanowire
and (101)-nanowire\label{fig2}}
\end{figure*}

In a basic DSSC operation, many processes take place influencing overall device
performances. These are the electron-hole generation upon visible light
absorption by the dye sensitizer, the charge carrier injection from the dye to
the conduction band (CB) of the oxide electrode, regeneration of the ground
state of the dye by a redox reaction through liquid iodide/triiodide
electrolytes.\cite{ORegan,Gratzel} Here one of the main concerns is the
reduction of photogenerated electron-hole recombination rate. Another point is
the binding strength of the sensitizer molecule to the oxide and the
reliability of this system in the ionic solvent. Therefore, the type of the dye
molecule and the surface properties of the semiconductor plays a key role in
the optimization of such processes.

Recent experiments have shown that one-dimensional nanostructures have several
advantages over nanoparticulate TiO$_2$ films.\cite{Boercker,KZhu} First of all
they have higher surface-to-volume ratios allowing many active sites to come
into contact with light harvesting molecules. Secondly, their one-dimensionality
forms a natural pathway for the charge carrier conduction from the point of
injection to the anode. Moreover, nanowires or nanotubes can exhibit a band
transport rather than a hoping mechanism between nanoparticles. Another crucial
aspect is that the electron-hole recombination rates were observed to be
ten times lower in nanotube-based films in comparison with films made from
nanoparticles.\cite{KZhu}

One of the key issues in DSSC is the type of the sensitizer dye and its 
interaction with the TiO$_2$ nanostructures. Commercially available Ru-based 
molecular complexes have shown up to 11\% solar energy conversion 
efficiencies.\cite{ORegan,Nazeeruddin1,Tachibana,Thompson,Nakade,Wang1,Benko,
Wang2,Nazeeruddin2} These metal-driven dyes can perform spatial charge 
separation and fast injection rates. Meanwhile, researchers have also focused on 
finding natural alternatives. Recently, antenna type novel organic 
donor-$\pi$-acceptor (D-$\pi$-A) dyes have been proposed as sensitizers to 
achieve charge transfer excitations reducing recombination rates that is the 
basic concern in simple skeleton light harvesting molecules. For instance, 
$\pi$-conjugated organic complexes with tetrahydroquinoline moiety as donor and 
a cyanoacrylic acid moiety as the acceptor were experimentally synthesized and 
theoretically studied.\cite{RChen1,RChen2,ORourke,Unal1} From this class, a 
highly efficient organic dye, TA-St-CA, contains a $\pi$-conjugated 
oligo-phenylenevinylene electron donor--acceptor moiety and a carboxyl group as 
anchoring group.\cite{Hwang,Lee,Zhang,Yang,Liang,Sharma,Mohammadi} Hwang 
\textit{et al.} achieved \% 9.1 photo-to-electric conversion efficiency with an 
open circuit voltage of 743 mV by designing a low cost TA-St-CA based 
DSSC.\cite{Hwang}

Natural dye pigments with a simple carbon skeleton structures such as cyanidin 
dyes are eco-friendly, widely available and cheap to produce. For instance, 
anthocyanin  (cyanidin-3-O-glucoside) can easily be extracted from 
plants.\cite{Senthil} The cyanidin family is well known and is proposed as an 
alternative to other dye 
sensitizers.\cite{Senthil,Tennakone,Cherepy,Ehret,Stintzing,Dai,Zheng,Kong, 
Galvano,Hao,McGhie,Parry, 
Sirimanne,Polo,He,Wongcharee,Duncan,Meng,Calogero1,Calzolari,Furukawa,Luo,Chang, 
Zhou,Buraidah,Aduloju,Calogero2,Chien} However, reports indicate the overall 
solar-to-electric energy conversion efficiencies below the current requirements. 
A deeper understanding of the bottlenecks in their DSSC applications is still 
needed.

In this study, we theoretically investigated the adsorption modes, electronic 
structures and absorption spectra of two different types of dyes on anatase 
TiO$_2$ nanowires having (001) and (101) facets. For this reason, we considered 
the D-$\pi$-A type organic complex TA-St-CA and the natural chromophore 
anthocyanin as dye sensitizers. We used both the standard and the screened 
exchange hybrid density functional theory (DFT) calculations to shed light on 
the main differences between the resulting nw+dye combined systems in vacuum as 
well as in solution treated with a modern nonlinear polarizable continuum model 
(PCM).

\section{Computational Details}

The DFT calculations have been performed using the projector-augmented wave 
(PAW) method\cite{Blochl} as implemented in the Vienna \textit{ab-initio} 
simulation package (VASP).\cite{Kresse1,Kresse2} Single particle electronic 
states have been expanded using plane wave basis sets up to a kinetic energy 
cutoff value of 400 eV. We used the standard generalized gradient approximation 
(GGA) to describe the exchange--correlation (XC) effects with the semicolon 
Perdew--Burke--Ernzerhof (PBE)\cite{Perdew} functional as well as the 
contemporary hybrid XC functional proposed by Heyd--Scuseria--Ernzerhof 
(HSE).\cite{Heyd1,Heyd2,Paier}  The latter is a nonlocal, screened Coulomb 
potential scheme with a range-separation. HSE functional largely heals the 
inherent bandgap underestimation of the standard DFT. To do so, 
HSE\cite{Heyd1,Heyd2,Paier} proposed to partially admix the exact Fock and the 
PBE exchange energies in the short range (SR) part as
\[
E_{\tiny\textbf{X}}^{\scriptsize\textrm{HSE}}=
a E_{\tiny\textbf{X}} ^{\scriptsize\textrm{HF,SR}}(\omega)+
(1-a)E_{\tiny\textbf{X}} ^{\scriptsize\textrm{PBE,SR}}(\omega)+
E_{\tiny\textbf{X}} ^{\scriptsize\textrm{PBE,LR}}(\omega)
\]
where $a$ is the mixing coefficient~\cite{Perdew2} and $\omega$ is the range 
separation parameter.\cite{Heyd1,Heyd2,Paier} Meanwhile, the long range (LR) 
part of the PBE exchange and the full PBE correlation energies are included in 
the HSE functional. In this way, the lack of proper self-interaction 
cancellation between the Hartree and exchange terms of the standard DFT is 
partly avoided leading to a substantial correction of the bandgap 
underestimation. In addition, the tendency of the standard DFT to give overly 
delocalized charge density distributions is also corrected to some extent. 
Therefore, this range separated hybrid density functional approach not only 
improves the bandgap related properties over the standard exchange--correlation 
(XC) schemes but also offers a better description of localized states such as 
the Ti $3d$ states of TiO$_2$ or isolated gap states caused by various impurity 
atoms.\cite{Janotti,Celik2} Hence, the HSE functional offers an improvement over 
PBE-calculated electronic and optical properties while the difference in the 
optimization of lattice structures is relatively less recognizable between PBE 
and HSE calculations. In fact, for dye+nanowire systems considered in this work, 
both XC functionals gave similar relaxed geometries and binding modes.

In order to obtain absorption spectra as the imaginary part of the dielectric 
function, $\varepsilon_2(\omega)$, from a density functional calculation, one 
considers the transitions from occupied to unoccupied states within the first 
Brillouin zone as the sum,
\begin{eqnarray*}
&& \hspace{-1cm} \varepsilon^{(2)}_{\alpha \beta}(\omega)=\frac{4 \pi^2 e^2}{\Omega}
\lim_{q\to 0}\frac{1}{q^2}\sum_{c,v,\mathbf{k}}
2\textsl{w}_{\mathbf{k}}\delta(\epsilon_{c\mathbf{k}}-\epsilon_{v\mathbf{k}}-\omega)
\nonumber \\
&& \hspace{2.4cm}\times\langle u_{c\mathbf{k}+\mathbf{e}_\alpha q}
\vert u_{v\mathbf{k}} \rangle\langle u_{c\mathbf{k}+\mathbf{e}_\beta q}
\vert u_{v\mathbf{k}} \rangle^*
\end{eqnarray*}
\noindent where the $c$ and $v$ show empty and filled states respectively, 
$u_{c\mathbf{k}}$ are the cell periodic part of the orbitals and 
$\textsl{w}_{\mathbf{k}}$ are the weight factors at each 
\textbf{k}-point.\cite{Gajdos}

The nw(001) and nw(101) nanowire models are constructed from the anatase 
TiO$_2$ bulk structures. They are considered in large tetragonal supercells 
with dye adsorbates such that the periodicity of the cell is chosen to be five 
times larger than one unit cell length along the nanowire axis to isolate the 
molecules from each other on the nanowires (as in Fig.~\ref{fig1} and 
Fig.~\ref{fig2}). In order to avoid unphysical interactions between the periodic 
images of the dye+nw structures, the supercells contain at least 20 {\AA} of 
vacuum separations along both of the lattice translation vectors perpendicular 
to the one along the nanowire axis. We fully optimized initial geometries by 
minimizing the Hellman--Feynmann forces on each ionic core to be less than 0.01 
eV/{\AA} based on the conjugate-gradients algorithm. None of the atoms were 
frozen to their bulk positions during these relaxation procedures. In this way, 
the bare nanowire models were previously shown to maintain the anatase structure 
without a major lattice distortion.\cite{Unal}

The effect of the solvent environment on the electronic structure of the 
dye--nanowire composed systems has been studied using the nonlinear 
polarizable continuum model (PCM) as implemented in the open-source code 
JDFTx.\cite{Gunceler,Sundararaman,Sundararaman2}  Since the nonlinear models 
incorporate the dielectric saturation effect, the nonlinear\cite{Gunceler} PCMs are 
expected to be more accurate than similar linear models in the presence of highly 
polar structures such as TiO$_2$.  In such models, the contribution to the dielectric 
function arising from the rotations of solvent molecules are modeled as a field of 
interacting dipoles, whose response function saturates with increasing external field. 
The cavity surrounding the solute is constructed self-consistently from its electron 
density, where the dielectric function of the solvent turns on at a critical electron 
density contour. The numerical value of the critical electron density  and the effective 
surface tension of the solute--solvent interface are highly solvent dependent. For 
solvents composed of small and highly polar molecules (such as water),  the effective 
surface tension at the interface is positive; whereas for solvents with large molecules 
and strong dispersion (van der Waals) interactions (such as chloroform)  this 
effective tension often has a negative sign. Additional technical details on the 
polarizable continuum model used as well as the numerical values for the critical 
electron densities and effective surface tensions of the solute--solvent interfaces can 
be found in the relevant publications.\cite{Gunceler,Gunceler2,Sundararaman2}

\section{Results \& Discussion}

The natural dye (cyanidin 3-O-glucoside) has been considered as a sensitizer
for DSSC applications by various groups.\cite{Tennakone,Cherepy,Ehret,Stintzing,
Dai,Zheng,Kong,Galvano,Hao,McGhie,Parry,Sirimanne,Polo,He,Wongcharee,Duncan,
Meng,Calogero1,Calzolari,Furukawa,Luo,Chang,Senthil,Zhou,Buraidah,Aduloju,
Calogero2,Chien} The main advantages over the other organic dyes have
been pointed out as the simple and low cost production together with its
wide availability in nature. Theory based studies on this class of dyes are
rather rare and mostly limited to isolated molecules in the gas phase.
We considered various probable initial adsorption configurations of
the cyanidin dye on both (001) and (101) anatase nanowire models. we carried
out the full optimization of the atomic coordinates using both PBE and HSE
functionals. Results indicate energetically favorable two different binding
modes on both nanowires as shown in Fig.~\ref{fig1}.~In order to differentiate
between these two modes we refer them as the physical and chemical bindings.
In the physical binding, the tail oxygen of the cyanidin part and the oxygen
of the OH group at the end of glucoside moiety interacts with two five-fold
coordinated surface Ti atoms (see Fig~\ref{fig1}a and Fig~\ref{fig1}c). In the
chemical binding mode, the OH group additionally loses its hydrogen to the
nearest surface oxygen site on both nw(001) and nw(101) models as shown in
Fig~\ref{fig1}b and Fig~\ref{fig1}d, respectively. Hence, the calculated
adsorption energies presented in Table~\ref{table1} indicate stronger binding
for the latter one.

The HSE-calculated Ti--O bond lengths between the cyanidin dye and nw(001)
are 2.04 {\AA} (on the cyanidin part) and 2.22 {\AA} (on the glucoside side)
for the physical adsorption mode. The second bond shortens to 1.88 {\AA} in
the chemical binding case. On nw(101), the corresponding bonds are 2.35 {\AA}
and 1.99 {\AA} for the physical binding while they are 1.97 {\AA} and
1.91 {\AA} for the chemical binding mode.

\begin{figure}[thb]
\hspace{4mm}
\includegraphics[width=7cm]{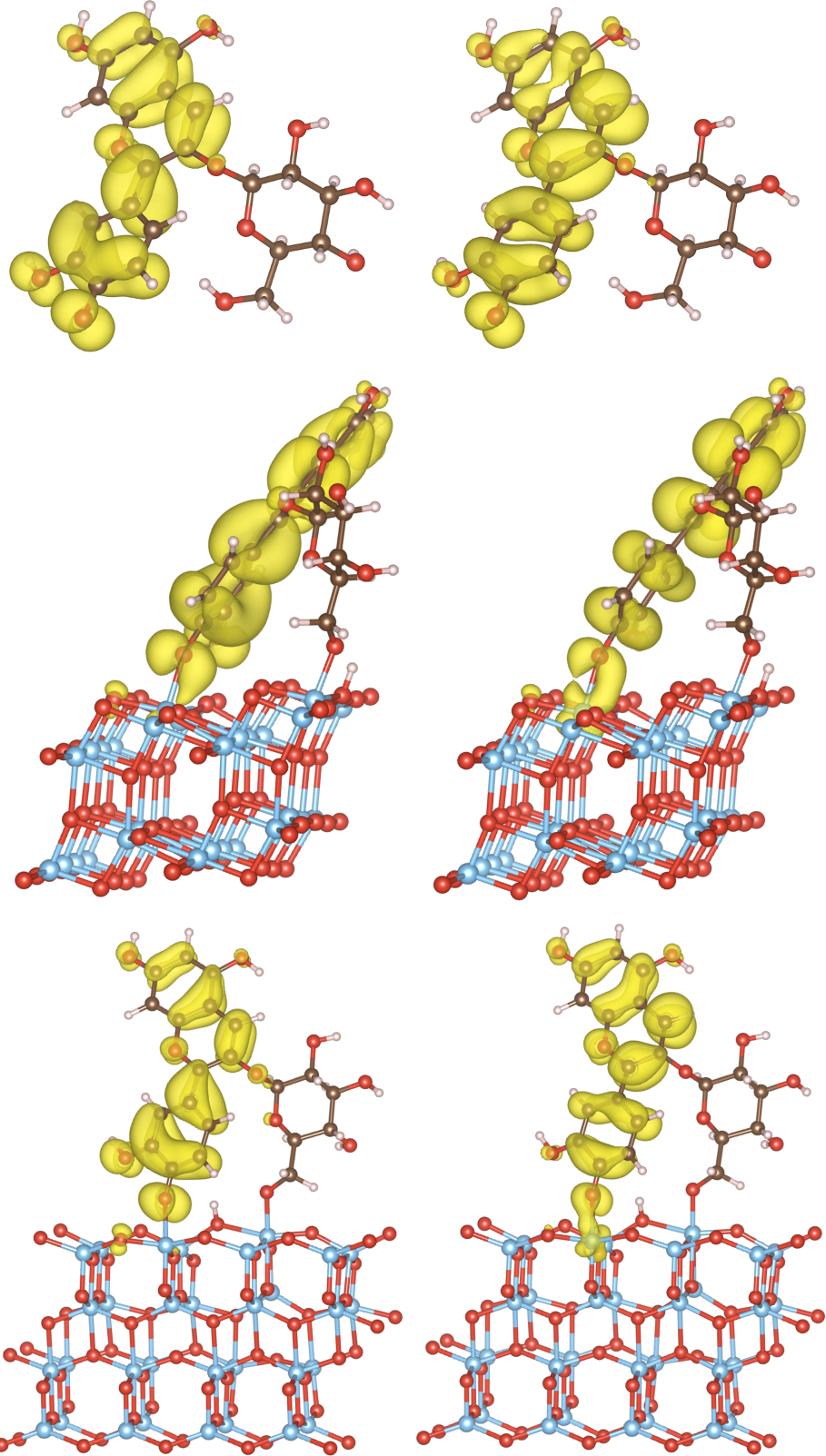}
\vspace{2mm}
\caption{Charge densities of the highest occupied states (on the left) and
the lowest unoccupied states (on the right) of cyanidin glucoside dye (top row),
 dye+nw(101) (middle row), and dye+nw(001) (bottom row).\label{fig3}}
\end{figure}

\begin{figure}[thb]
\hspace{4mm}
\includegraphics[width=7.3cm]{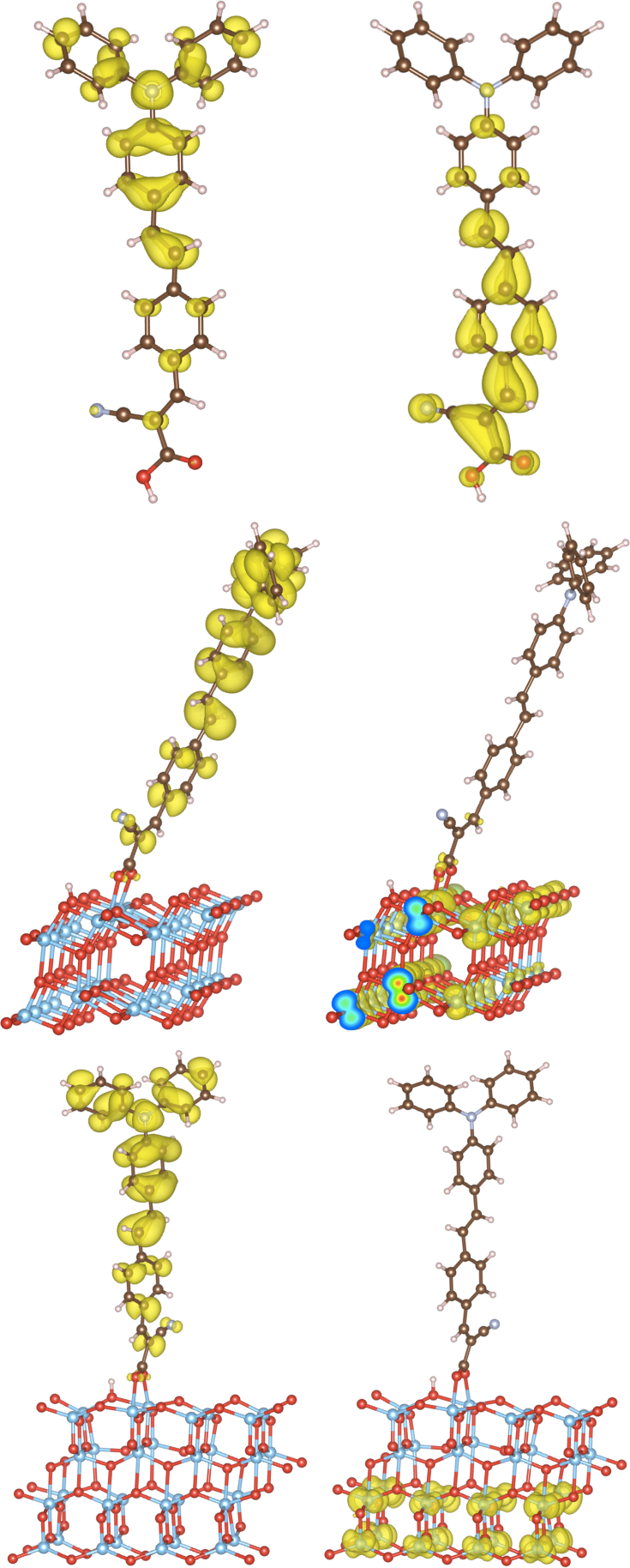}
\vspace{2mm}
\caption{Charge densities of the highest occupied states (on the left) and
the lowest unoccupied states (on the right) of TA-St-CA dye (top row),
 dye+nw(101) (middle row), and dye+nw(001) (bottom row).\label{fig4}}
\end{figure}

The geometry optimizations using the standard PBE and modern HSE functionals
gave similar final structures except the chemical binding mode of cyanidin
molecule on the (101) nanowire. PBE calculations predict much stronger
interaction between the natural dye and the (101) surface of the oxide.
The PBE functional leads to considerable local distortion around the adsorption
site where the bond lengths abruptly change. For instance, PBE-relaxation
breaks the bond between the surface Ti which the dye is anchored at and the
adjacent surface O which captures the hydrogen from the dye. The separation
between them increases from the typical surface O--Ti bond length of 1.97 {\AA}
to 3.81 {\AA}. While PBE causes such a local reconstruction, HSE calculations
yield an adsorption geometry where the (101) nanowire keeps its bare surface
structure. Therefore, calculated binding energies for this case are 2.35 eV and
1.20 eV with PBE and HSE schemes, respectively (see Table~\ref{table1}).
Experimental studies does not report a binding at the chemisorption level  to
support the PBE predictions for this specific case. Therefore, this can be seen
as one of the examples where the standard exchange--correlation functionals end
up with peculiar results.

The TA-St-CA molecule on the (001) and (101) facets of the anatase nanowires
has two different low energy adsorption structures, the monodentate and the
bidentate binding modes as shown in Fig~\ref{fig2}. The monodentate binding
portrays perpendicular alignment with respect to the nanowire axis as a result
of the single bond formation between the tail oxygen and surface Ti atom. The
bidentate mode is similar to the cyandin case because of the additional H
transfer from the OH group at the tail to the nearest surface oxygen site on
both of the nanowire types. The loss of hydrogen from the dye to the surface
enables another O--Ti bond formation between the molecule and the oxide.

In the bidentate case, both O-Ti bonds are 2.01 {\AA} between TA-St-CA and
nw(101). Likewise, they are 2.03 {\AA} on nw(001). Monodentate TA-St-CA forms
a single bond of 2.14 {\AA} on nw(101) while it is 2.36 {\AA} on nw(001).
These results expectedly indicate stronger binding in favor of the bidentate
adsorption. A similar conclusion can be drawn from the calculated binding
energies in Table~\ref{table1}.

\begin{table*}[bth]
\caption{Calculated adsorption energies, $E_{\rm ads}$, of the dye--nanowire 
systems\label{table1}}
\begin{ruledtabular}
\begin{tabular}{cccccccccc}
& \multicolumn{4}{c}{@(001)} &&
\multicolumn{4}{c}{@(101)}\\ \cline{2-5} \cline{6-10}
Dye & PBE & HSE & PBE+PCM$^1$ & PBE+PCM$^2$ &&
PBE & HSE & PBE+PCM$^1$ & PBE+PCM$^2$ \\[1mm] \hline
cyanidin glucoside (physical) & $-1.35$ & $-1.22$ & $-1.13$ & $-0.78$&&
$-1.47$ & $-1.42$ & $-1.16$ & $-0.74$\\ \hline
cyanidin glucoside (chemical) & $-1.67$ & $-1.30$ & $-1.33$ & $-0.89$&&
$-2.59$ & $-1.20$ & $-2.08$ & $-1.44$\\ \hline
TA-St-CA (monodentate) & $-0.24$ & $-0.28$ & $-0.03$ & ~~$0.18$&&
$-0.74$ & $-0.72$ & $-0.50$ & $-0.26$\\ \hline
TA-St-CA (bidentate) &$-1.05$ & $-1.02$ & $-0.81$ & $-0.58$&&
$-0.84$ & $-0.93$ & $-0.64$ & $-0.43$\\
\end{tabular}
\end{ruledtabular}
\begin{flushleft}
$^1$ in CH$_3$Cl using nonlinear PCM \\
$^2$ in H$_2$O using nonlinear PCM
\end{flushleft}
\end{table*}

Molecular complexes with anchoring groups are rather obvious to yield larger
adsorption. However, the main difference between TA-St-CA and cyanidin dyes
is not their sizes. The most important factor is that while TA-St-CA is
specifically designed to do an intramolecular charge separation from the
antenna part to the acceptor moiety which attaches to the oxide surface, both 
the HOMO and LUMO charge densities are localized on the cyanidin part of the
cyanidin-3-O-gloucoside. The glucoside group does not give any contribution
to the frontier molecular orbitals in the case of cyanidin dye. This can be
seen from the calculated charge density distributions for the corresponding
states of the molecules in their gas phase as shown at the first rows of 
Fig~\ref{fig3} and Fig.~\ref{fig4}.

\begin{figure*}[htb]
\epsfig{file=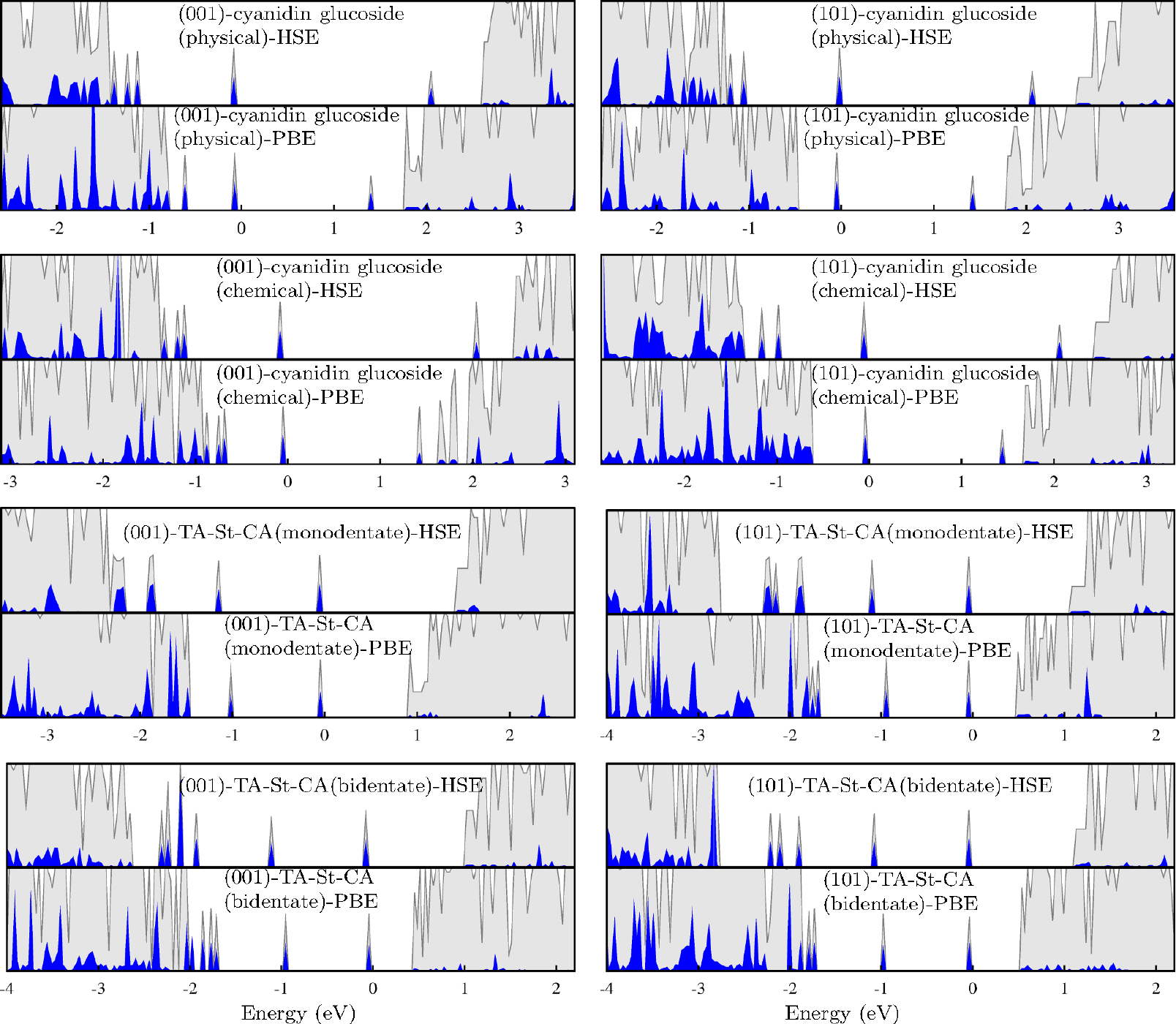,width=16cm}
\caption{Densities of states for the dye+nanowire systems (in arbitrary units) 
calculated using the PBE and HSE functionals. The partial DOS contributions of 
the dye molecules are shown as dark (blue) shades.\label{fig5}}
\end{figure*}

In fact, the comparison of the charge density redistribution features for
the lowest lying optical excitation becomes more important when these
molecules are attached to the oxide surfaces. Therefore, the spatial charge
densities of the highest occupied and the lowest unoccupied states of dye+nw
combined systems have been calculated to discuss the charge injection features
of these two different types of dyes. The HSE results are presented in 
the middle and at the bottom rows of Fig.~\ref{fig3} and Fig.~\ref{fig4}  
for the nw(101) and  the nw(001) cases, respectively. The charge densities of the 
frontier molecular orbitals of the cyanidin dye remain very similar to their gas 
phase distributions even if it forms two O-Ti bonds with both the anatase (001) 
and (101) surfaces. These two states also appear in the band gap of nanowires 
as well-localized isolated states, one being filled and the other one being 
empty (see Fig.~\ref{fig5}). Hence, the lowest vertical excitation does not 
involve a charge injection to the CB of the semiconductor. Such an excitation is 
prone to yield an electron--hole recombination. Therefore, HSE-calculated charge 
density results might partially explain why this type of natural dye pigments 
end up with relatively low incident photon to current efficiencies (IPCE).

The charge density distributions of the frontier orbitals of an isolated TA-St-CA 
molecule in vacuum have been investigated by Zhang\textit{et al.}\cite{Zhang} 
and Mohammadi \textit{et al.}\cite{Mohammadi} using
hybrid DFT calculations. Their gas phase results agree with our calculations
for the isolated molecule case as shown in the top row of Fig.~\ref{fig4}. 
In the case of TA-St-CA on TiO$_2$, the intramolecular charge transfer character 
from the triphenylamine part as donor to the cyanoacrylic acid group as the
acceptor seems to be significantly altered after the molecule is adsorbed
on the anatase nanowires. The lowest lying excitation involve a transition 
from the HOMO-like state which is spatially well-localized on the antenna
moiety to the CB of the anatase as shown for the TA-St-CA+nw combined 
systems in Fig.~\ref{fig4}. Therefore, our calculations indicate that upon
adsorption the loss of donor-$\pi$-acceptor gas phase feature of TA-St-CA
changes in favor of better charge injection into the CB of the oxide.
This might also be interpreted as accounting for the reduction of
recombination rates in the case of phenylenevinylene-conjugated
D-$\pi$-A type sensitizers.

The adsorption energies of the molecules have been calculated using
\[
 E_{\rm ads}=E_{\rm dye+nw}-(E_{\rm dye}+E_{\rm nw})
\]
where $E_{\rm dye+nw}$, $E_{\rm dye}$, and $E_{\rm nw}$ are the relaxed 
supercell energies of dye adsorbed nanowire, the isolated dye molecule, and the 
bare anatase nanowire, respectively. Both the PBE and HSE functionals were used
for the vacuum calculations as presented in Table~\ref{table1}. Then, we 
obtained adsorption energies in solution for chloroform and water using a new 
non-linear PCM\cite{Gunceler} with the PBE functional. Our tests show that 
similar conclusions can be drawn when the solvent effects are included within 
the hybrid HSE scheme. The vacuum results with PBE and HSE functionals are only 
slightly different from each other except the chemical binding of the cyanidin 
dye on the nw(101). For this case, the energy difference of 1.39 eV is due to 
the fact that the PBE functional overestimates the bonding between the hydrogen 
atom captured from the dye with the nearest surface oxygen, which results in an 
additional local modification of atomic positions relative to the HSE-optimized 
structure where surface Ti-O row seems to be disturbed less.

As expected, single bond formation leads to a weaker adsorption energy for each 
type of dye. The loss of H from the OH group of the dye to the nearest surface 
oxygen causes the formation of a second bond. This situation enhances the 
binding appreciably, especially in the bidentate mode of TA-St-CA. Clearly, the 
binding of the molecules appears to be noticeably stronger in vacuum. As a 
nonpolar solvent with a dielectric constant of 4.8, chloroform has a little 
effect on the adsorption energies. However, water, being a polar solvent, 
weakens the bond(s) between the dye and the nanowire, considerably. For 
instance, the results show that monodentate TA-St-CA is washed away from the 
surface. In the other cases, the dye molecules can keep moderate binding with 
the oxide even in water environment. 

We note that the binding energies change in the positive direction as seen in 
Table~\ref{table1}. The reason is that the binding sites on the TiO$_2$ nanowire 
and the dye molecule interact strongly with the solvent. On the other 
hand, such an interaction is absent in a vacuum calculation leading to a more 
negative binding energy.  When we check the differences between the calculated 
Kohn-Sham eigenvalues in vacuum and in solution, we see that the energy levels 
are slightly shifted with respect to each other. Since H$_2$O is a more polar 
solvent than CHCl$_3$ with a higher dielectric constant, it interacts more 
strongly with the binding sites; we therefore see that binding energies in 
H$_2$O are more positive than the binding energies in CHCl$_3$.

Since the dye molecules develop a bonding interaction with the oxide 
surface, dye-related HOMO-like levels appear in the band gap of the oxide as 
isolated and well-localized states. In addition, their LUMO levels show a 
strong dispersion inside the CB of TiO$_2$. Those can no longer be 
considered as molecular energy levels. Therefore an optical excitation starts 
from the HOMO-like dye-related state to the states in the CB.

For an efficient DSSC, the lowest lying absorption peaks (in the visible range)
mainly involve transitions from the HOMO-like dye-related states which appear
above the VB to the states in the CB of TiO$_2$. If the dye molecules had
broken apart and been dissolved in solution one could have expected the loss of
main features of the absorption peaks. However, we have shown that the water or
chloroform can only weaken the bonds between the molecules and TiO$_2$. 
When we check the differences between the calculated Kohn--Sham eigenvalues 
in vacuum and in solution, we see that the single particle energy levels are 
slightly shifted with respect to each other. Therefore, from a theoretical 
point of view, the difference in the calculated electronic properties between 
the standard PBE and modern HSE exchange--correlation schemes is much larger 
than the energy shifts due to PCM.

\begin{figure}[htb]
\includegraphics[width=8.6cm]{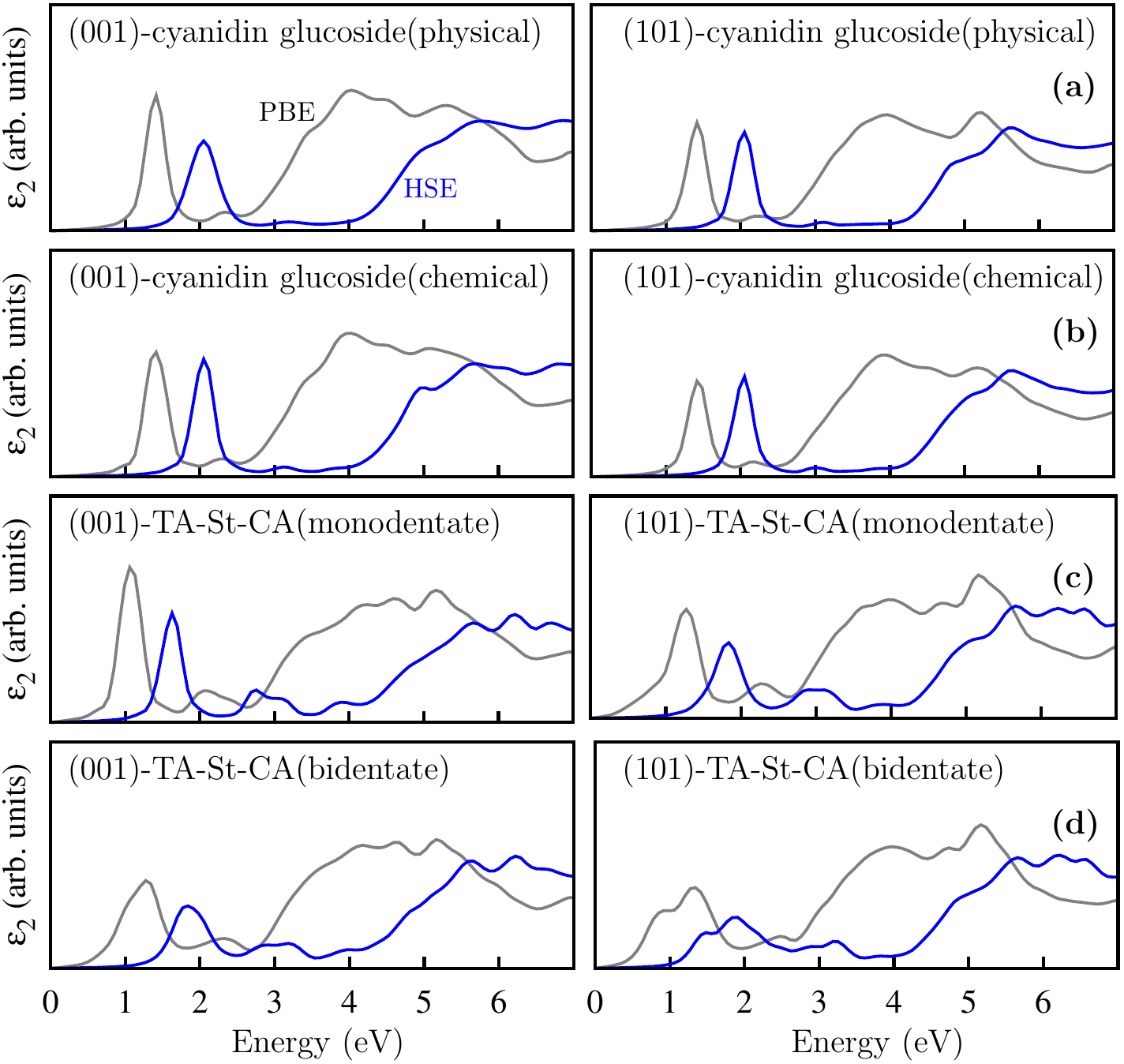}
\caption{First two rows display the calculated absorption spectra of the 
dye+nanowire combined systems : (a) physical (b) chemical binding of 
cyanidin glucoside on the anatase nw(001) and nw(101), respectively. The last 
two rows show the absorption spectra for (c) monodentate TA-St-CA and (d) 
bidentate TA-St-CA on anatase nw(001) and nw(101), respectively. \label{fig6}}
\end{figure}

In order to discuss the electronic structures of the dye+nanowire combined 
systems the partial density of states has been calculated for each binding mode
using both the PBE and HSE functionals as presented in Fig.~\ref{fig5}.
Since, the DOS structures calculated using the PBE and HSE functionals were 
aligned with respect to their deep core states, the HSE VB edge lies lower than 
the PBE one. Therefore, in the case of HSE, several new dye-related states fall 
in the band gap, which seem to occupy the top of the VB in the PBE DOS 
structures. The inclusion of partial short ranged exact exchange 
corrects the description of Ti $3d$ states reasonably. Since the CB of
TiO$_2$ is composed of these $d$ bands, the CB gets shifted up to higher 
energies. Therefore, HSE functional heals the band gap underestimation of 
the standard exchange-correlation schemes not only by shifting up of the CB 
edge but also by lowering the VB edge. Meanwhile, the presence of new dye 
related gap states cause a significant narrowing of the band gap, which
functionalizes the combined system to be active in the visible region.

The absorption spectra of the dye+nanowire systems have been obtained at the 
PBE and HSE levels by calculating the dipole transition matrix elements between
the occupied and the empty states. Similar computations were performed 
previously for the bare anatase nanowires.\cite{Unal1} The presence of new 
isolated and occupied gap states originating from the dye adsorbates just above 
the VB edge of the oxide is desirable for functionalization of TiO$_2$ in the 
visible part of the spectrum. When one compares the calculated optical spectra, 
similarities can be found between the two nanowire types as well as between the 
results of PBE and HSE functionals as shown in Fig.~\ref{fig6}. Among these 
characteristics, the first absorption peaks in each case are associated with the 
vertical transitions from the dye-related HOMO-like gap states to the lowest 
lying unoccupied states. Those sharp peaks indicate a significant red shift of 
the absorption threshold into the visible region which is favorable for titania 
based photovoltaic applications. Although the overall features look alike, 
PBE-calculated spectra is also considerably red shifted relative to those of the 
HSE due to the local density approximation (LDA) to exchange-correlation effects 
giving rise to an underestimation of the band gap of TiO$_2$. Moreover, the red 
shifting of the PBE spectra with respect to the HSE results gets even larger for 
the higher lying transitions. 

Experimentally, Senthil \textit{et al.}\cite{Senthil} reported the first absorption 
peak of cyanidin-3-O-glucoside adsorbed on TiO$_2$ nanoparticles at 2.22 eV. 
Although the efficiency they have obtained with this dye is still below the 
requirements as they said, the main reason remained unclear.  In our calculation 
for the cyanidin glucoside on both of the nanowire types, the first 
absorption peak positions are found around 1.5 eV and 2.2 eV with PBE and HSE 
functionals, respectively, which coincide with the energy difference between the 
isolated dye-related gap states as seen in the top two panels of 
Fig.~\ref{fig5}. Therefore, the lowest lying transitions in each case 
essentially involve an intramolecular excitation from the HOMO-like state to the 
LUMO-like state in the band gap as seen in first two panels of Fig.~\ref{fig5}.  
The corresponding charge density distributions can also be seen in 
Fig.~\ref{fig3}. Although the first absorption peak seems to drive the 
photoresponse of the combined system to visible region, the associated 
excitation takes place on the dye itself without any sign of charge injection 
into the semiconductor. This might be seen as one of the factors why these type 
of natural dyes have relatively low incident photon to current efficiency (IPCE) 
observed in experiments.\cite{Senthil} 

Recently, Hwang \textit{et al.} reported an overall solar-to-energy conversion 
efficiency of 9.1 \% with TA-St-CA sensitizer.\cite{Hwang} The experimental
optical absorption peaks were found around 2.4 eV.\cite{Hwang,Lee} This
value agrees with our HSE results for the bidentate TA-St-CA. The remarkable
difference in the efficiencies between the natural cyanidin and organic TA-St-CA
sensitizers needs an understanding from a theoretical perspective. First of all, the 
adsorption of TA-St-CA causes the LUMO to resonate with the Ti $3d$ 
states and get delocalized in the CB of the anatase nanowires. Therefore, the first 
peaks corresponding to absorption in the visible region are due to transition 
from the dye-related HOMO-like gap states to the states in the CB as seen in the 
bottom two panels of Fig.~\ref{fig6}.  The inclusion of screened exact exchange 
in the HSE functional shifts the CB edge to higher energies such that the
gap values become $\sim1$.3 eV whereas the PBE gaps are considerably 
underestimated as seen in Fig.~\ref{fig5}. Consequently,  the first peak 
positions agree with the energy difference of those states involved in the 
transition. When we look at the corresponding charge density plots in 
Fig.~\ref{fig4}, we see a charge injection into the oxide through electron-hole 
generation upon visible light absorption. In comparison to the cyanidin 
case, the appearance of a group of dye-related states just above the VB edge 
in the HSE results translates into an absorption ability of the 
TA-St-CA+nanowire system in a wider range of the visible spectrum with varying 
oscillator strengths. For the TA-St-CA sensitizer, transitions might arise
from these group of gap states below the HOMO-like level to the states in the 
CB causing shallow peaks around ~3 eV shown in Fig.~\ref{fig6}c and d. 
The results indicate that the cyanidin dye is not successful in sensitizing 
anatase nanowires in the region between the first peaks and the main body of 
the higher energy contributions. This is much more pronounced in the spectra 
obtained with HSE functional. Higher frequency contributions to absorption in 
the UV part of the spectrum are mostly associated with the interband transitions 
from the states in the VB to the states in the CB of the dye+nanowire system. 
Both molecules bring the lowest lying peak which drops the absorption threshold 
into the visible region as seen in Fig.~\ref{fig6} for both nanowire types. The 
TA-St-CA+nanowire combined system has more favorable optical properties than 
the cyanidin glucoside+nanowire structure.

\section{Conclusions}

The natural chromophore cyanidin-3-O-glucoside and the D-$\pi$-A type organic 
dye TA-St-CA have been separately considered on anatase TiO$_2$ nanowires having
either (101) or (001) facets. The standard and the screened Coulomb hybrid 
density functional theory calculations were performed to understand the 
adsorption modes, electronic structures and absorption spectra of dye+nanowire
combined systems. The binding energies have been obtained in vacuum and in 
solution using a new non-linear PCM. Our results indicate a significant band gap
narrowing upon adsorption on both of the nanowire facets due to the appearance
of a number of isolated dye-related gap states. Their number increases when
a second bond is formed between the dye molecule and the oxide. The 
HSE-calculated VB edge lies energetically lower with respect to that 
obtained using the PBE functional. This causes several dye-related states to 
fall in the band gap just above VB. Therefore, the HSE functional gives a 
larger number of occupied gap states. In addition, the HSE functional corrects 
the band gap underestimation of the standard approximations to the 
exchange-correlation energy by admixing partial screened exact exchange. 
Therefore, unoccupied Ti $3d$ states are better described with the HSE 
functional leading to a significant shift of the CB up to higher energies.

The cyanidin dye can form single and double bonds through its tail oxygens with 
the surface Ti ions on anatase. The latter is a chemical binding where 
the nearest surface oxygen captures an H from the tail OH group of cyanidin dye. 
Both adsorption modes bring several new occupied states above the VB and an 
empty state below the CB of TiO$_2$ at both the PBE and the HSE levels of 
theory. The lowest lying transition starting from the HOMO-like dye-related gap 
state to the empty LUMO-like state below the CB takes place on the dye itself 
giving a sharp absorption peak in the visible range.This can be understood as a 
factor limiting a subsequent charge injection into the nanowire and might end up 
with recombination of photoexcited electron-hole pairs. TA-St-CA shows mono- and 
bidentate binding modes on both of the anatase nanowires. When adsorbed, its 
intramolecular charge transfer character gets favorably modified toward a charge 
injection into the oxide. This might help reduce the recombination rates of 
charge carriers. The isolated filled gap states originating from the bidentate 
binding of TA-St-CA sensitizer significantly narrow the energy gap and 
effectively functionalize the anatase nanowires to actively absorb a broader
range of the visible spectrum. Inclusion of nonlinear solvation effects 
indicates dissociation of the monodentate TA-St-CA from the nanowires in 
water where the bidentate TA-St-CA develops a moderate binding. In a more polar 
ionic solution, TA-St-CA needs a firmer anchoring to the oxide.
Although the cyanidin molecule has a strong binding on the (101) and 
(001) nanowires even in an electrolyte, the absorption properties are weaker
and might suffer from recombination of photogenerated electron-hole pairs.
Bidentate TA-St-CA+nanowire systems can achieve directional charge transfer 
excitation to increase charge injection probabilities, allow absorption in the 
wide range of visible spectrum with enhanced light harvesting, and 
exhibit moderate binding in solution required to reduce degradation of 
possible device operation.

%\section{Acknowledgments}
\begin{acknowledgments}
This work was supported by T\"{U}B\.{I}TAK, The Scientific and Technological
Research Council of Turkey (Grant \#110T394). Computational resources were
provided by ULAKB\.{I}M, Turkish Academic Network and Information Center.
\end{acknowledgments}

\end{document}